\documentclass[10pt]{article}
\newcommand{\be}{\begin{equation}}
\newcommand{\bea}{\begin{eqnarray} \nonumber}
\newcommand{\ee}{\end{equation}}
\newcommand{\eea}{\end{eqnarray}}
\newcommand {\lan} {\langle}
\newcommand {\ran} {\rangle}
 \def\(({\left(}
 \def\)){\right)}
\def\[[{\left[}
\def\]]{\right]}
\def\bi{\bibitem}
\def \form#1 {eq. (\ref{#1}) }
\def \parziale#1#2  {{\partial {#1} \over \partial {#2}}}

\def \ba#1 {\overline{#1}}

\def  \eps{\epsilon}

\begin{document}
\title{Local overlaps, heterogeneities \\ and \\the local fluctuation dissipation relations}
\author{ Giorgio Parisi \\
  Dipartimento di Fisica, INFM, SMC and INFN, \\
Universit\`a di Roma {\em La Sapienza}, P. A. Moro 2, 00185 Rome, Italy. }
\maketitle
\begin{abstract}

In this paper I introduce the probability distribution of the local overlap in spin glasses.  The properties of the
local overlaps are studied in details.  These quantities are related to the recently proposed local version of the
fluctuation dissipation relations: using the general principle of stochastic stability these local fluctuation
dissipation relations can be proved in a way that is very similar to the usual proof of the fluctuation dissipation
relations for intensive quantities.  The local overlap and its probability distribution play a crucial role in this
proof.  Similar arguments can be used to prove that all sites in an aging experiment stay at the same effective
temperature at the same time.
 \end{abstract}
\section{Introduction}
Up to now in disordered systems the overlap and its distribution were considered as global quantities, that were defined
for the whole system \cite{RM}.  However in systems with quenched disorder it is possible to define (in a non-trivial way) a local
 overlap that has a point-dependent probability distribution \cite{I}.  This new object has remarkable properties that we
explore in this paper.

One of the most interesting results is related to the local generalization of fluctuation dissipation relations (FDR)
in off-equilibrium dynamics. 

It is well known that the fluctuation dissipation relations in off-equilibrium dynamics are a crucial tool to explore
the landscape of a disordered system \cite{CUKU,framez,CKPA,revaging}.  These fluctuation dissipation relations are
different from the prediction of the fluctuation dissipation theorem at equilibrium.  They can be expressed in a rather
simple form that can be easily interpreted from the theoretical point of view.  Moreover the main parameter entering in
the fluctuation dissipation relations has simple interpretation from the point of view of equilibrium statistical
mechanics \cite{GUERRA,AI,SOL,NOI}.  This fact has consequence that for a given system the form of the FDR is universal
in off-equilibrium dynamics, i.e. it does not depends on the details of the dynamics and on the way in which the system
is put in an off-equilibrium situation (as soon the system remains slightly out of equilibrium).

In the mostly studied case one considers observables that are the average over the whole sample \cite{FR,RM,P,K,IG,BC,HO}. 
In this case the static-dynamic relations connect the FDR to the static average of global quantities.  Recent there have
been a few investigations on FDR that involve only given local variables \cite{CC,MR}.  It turns out that using the
probability distribution of the single spin overlap it is possible to give a theoretical foundation to these local FDR
and to derive the appropriate static dynamic relations, that involve the probability distribution of the local overlap
\cite{I}.

The paper is organized as follows: in the second section we define our main new theoretical tool: the probability
distribution of the local overlap; we compute its properties in a few simple cases, in the third section we recall the
usual global FDR, while in the next section we recall the proposed local FDR \cite{CC,MR,I}.  In the fifth section we
derive from general principle the local FDR, we prove the appropriate static-dynamic relations and show that in an
ageing regime, in spite of the existence of local heterogeneities all sites at a given time must be characterized by the
same effective temperatures.  In the next section, before the conclusions, we discuss some methods to compute the
probability distribution of the local overlap.  Finally, it the appendix, we present some consideration on systems with
finite volume.

\section{The local overlap distribution}
\subsection{Some heuristic considerations}
This section contains the definition of the local overlap probability distribution $P_{i}(q)$.

Usually the overlap of two equilibrium configurations ($\sigma$ and $\tau$) is defined as
\be
q={\sum_{i}q_{i} \over N} \ ,
\ee
where $q_{i}=\sigma_{i}\tau_{i}$.
For a given sample we can define the overlap probability distribution $P_{J}(q)$, where $J$ label the sample.  In the
glassy phase the function $P_{J}(q)$ depends on the sample also for very large samples: the global overlap probability
distribution $P_{J}(q)$ is not a self-averaging quantity.  The physically interesting quantity is
\be
P(q)=\overline{P_{J}(q)}\ ,
\ee
where the bar denotes the average over the couplings $J$. 

In a different approach \cite{NOI} one considers the response of the system to the appropriate perturbation and in this
way one can define for a given sample a function $P_{r}(q)$, that for large systems should be self-averaging, i.e.
$J$-independent.  This new order parameter distribution codes the thermodynamic responses of the system to random
perturbations.  According to the principle of stochastic stability \cite{GUERRA,AI,SOL,NOI} it should coincide with
$P(q)$.

It is evident that the definition of the local overlap must be rather different from that of the global overlap.  Indeed in a naive
approach the local overlap of two equilibrium configurations (i.e. $\sigma_{i}\tau_{i}$) is always equal to $\pm 1$;
if use a naive definition the probability distribution of the local overlap should be the sum of two delta functions at
$\pm 1$: in this way one gets a trivial result.

Moreover our aim is to define a site dependent, sample dependent $P_{i}(q)$ such that
\begin{equation}
P_{g}(q)=\prod_{i=1,N}\left( P_{i}(q_{i})dq_{i}\right) \delta\left( {\sum_{i}q_{i} \over N} -q\right)
\end{equation}
is a self-averaging quantity.  In other words $P_{g}(q)$ cannot coincide with $P_{J}(q)$ but it should be equal to 
$P_{r}(q)$. In the following we will extend the approach of  \cite{NOI} in order to arrive to a definition of $P_{i}(q)$
that satisfies the aforementioned properties.

\subsection{The definition of the local overlap}
Let us start from a spin glass sample and let us  consider $M$
identical copies  of our sample: we introduce $N \times M$ $\sigma^{a}_{i} $ variables where $a=1,M$
(eventually we send $M$ to infinity) and $N$ is the (large) size of our sample ($i=1,N$).  The Hamiltonian in this Gibbs
ensemble is just given by
\begin{equation}
H_{K}(\sigma)=
\sum_{a=1,M} H(\sigma^{a})  +\eps H_{R}[\sigma] \ , 
\end{equation}
where $H(\sigma^{a})$ is the Hamiltonian for a fixed choice of the couplings and the  $H_{R}[\sigma]$
is a random Hamiltonian that couples the different copies of the system.  A possible choice is 
\begin{equation}
    H_{R}[\sigma]=\sum_{a=1,M;i=1,N}K^{a}_{i}\sigma_{i}^{a}\sigma_{i}^{a+1} \ , \label{PS}
\end{equation}
where the variables $K^{a}_{i}$ are identically distributed independent random Gaussian variables with zero average and
variance 1.  In this way, if the original system was $d$ dimensional, the new system has $d+1$ dimensional, where the
planes are randomly coupled. We can consider other ways to couple the systems (e.g. $
H_{R}[\sigma]=\sum_{a,b=1,M;i=1,N}K^{a,b}_{i}\sigma_{i}^{a}\sigma_{i}^{b}$). An other possibility is
\be
H_{R}[\sigma]=\sum_{a,b=1,M;i,j=1,N}K^{a,b}_{i,j}\sigma_{i}^{a}\sigma_{j}^{b} \label{SS}
\ee
where the  variables $K$ are identically distributed independent random Gaussian variables with zero average and
variance $(NM)^{-1}$. As we shall see later the form of $H_{R}$ is not important: its task it to weakly couple the 
different planes that correspond to different copies of our original system. The first choice \form{PS} is the simplest 
to visualize and it is the fastest for computer simulations, the last choice \form{SS} is the simplest one 
to analyze from the theoretical point of view. In the following we do not need to assume a particular choice.

Our central hypothesis is that all intensive self average quantities are smooth function of $\eps$ for small $\eps$. 
This hypothesis is a kind of generalization of stochastic stability.  According to this hypothesis the dynamical local correlation
functions and the response functions will go uniformly in time to the values they have  at $\eps=0$.

We now consider in the case of non-zero $\eps$ two equilibrium configurations $\sigma$ and $\tau$ 
and let us define for given $K$ the site dependent overlap 
\begin{equation}
q_{i}(\sigma.\tau)={\sum_{a=1,M} \sigma^{a}_{i} \tau^{a}_{i} \over M}
\end{equation}
We define the $K$ dependent probability distribution  $P^{K}_{i}(q)$ as the probability  
distribution of the previous overlap. If we average over $K$ at fixed $\eps$ we can define
\begin{equation}
P^{\eps}_{i}(q)=\overline{P^{K}_{i}(q)} \ ,
\end{equation}
where the bar denotes the average over $K$.
We finally define
\begin{equation}
P_{i}(q)=\lim_{\eps \to 0}P^{\eps}_{i}(q) \  ,
\end{equation}
where the limit $\eps \to 0$ is done {\sl after} the limits $M \to \infty$ and $N \to \infty$ 
(alternatively we keep $\eps M$ and $\eps N$ much larger than 1).

In order to be consistent we the usual approach we should have that if define
\begin{equation}
q_{t}={\sum_{i=1,N}q_{i} \over N} \ ,
\end{equation}
the probability distribution $ P_{t}(q) $ of $q_{t}$ should be self-averaging (i.e. $J$ independent and it should
coincide with the function $P(q)$ that is the average over $J$ of $P_{J}(q)$:
\begin{equation}
P_{t}(q)=P_{g}(q)\equiv E_{J}\[[ P_{J}(q) \]] \label{CON} \ .
\end{equation}
This crucial relation will be proved in the next section.

In the nutshell the construction is rather simple.  We consider $M$ weakly coupled copies of the original lattice and in
the limit $M$ going to infinity we can define {\sl local} thermodynamics averages.  We will assume that we stay also in
the infinite volume limit.  A discussion of what happens for finite $N$ but $M=\infty$ will be presented in the
appendix.

\subsection{Some heuristic considerations}
Which is the rational of this baroque construction?  The heuristic idea is simple.  For finite $N$ and given $J$ we can
approximatively decompose the usual Boltzmann expectation value $\lan \cdot \ran$ into different states labeled by
$\alpha$:
\begin{equation}
\lan \cdot \ran = \sum_{\alpha}w_{\alpha} \lan \cdot \ran_{\alpha}
\end{equation}
If we define the state dependent magnetizations 
\begin{equation}
m^{\alpha}_{i}=\lan \sigma_{i} \ran_{\alpha} \ .
\end{equation}
we have that the overlap among two states is
\begin{equation}
q^{\alpha,\gamma}={\sum_{i=1,N}m^{\alpha}_{i}m^{\gamma}_{i} \over N}
\end{equation}
and that the usual overlap distribution is given by
\begin{equation}
P_{J}(q)= \sum_{\alpha,\gamma} w_{\alpha}w_{\gamma}\delta(q^{\alpha,\gamma}-q) \ .
\end{equation}
The average over $J$ gives the usual $P(q)$.

However we do not want to average over the $J$;we want to stick to a given system.  In order to make a sensible
selfaveraging definition we have to consider an ensemble of system so we will consider the original system plus a small
random perturbation $H^{R}$.  However, if the expectation value of the random perturbation in the various states
($E^{R}_{\alpha}\equiv \lan H^{R}\ran_{'\alpha}$) is much larger than one, the new $w$'s will be proportional to
\begin{equation}
w_{\alpha}\exp (-\beta E^{R}_{\alpha})
\end{equation}
and the states will be completely reshuffled.  In a first approximation we can assume that the set of states remains the
same after the perturbation, but the thermodynamically most relevant states do change.  The way in which they change
depends on the distribution of the $\alpha$.  However the principle of stochastic stability tell us the we get the {\sl
same} function $P(J)$ if we do the average over the random perturbation at fixed $J$ or over the ensemble of the $J$ in
absence of the random perturbation.

At non zero $\eps$ we can assume that the systems we state is same state $A$ with a probability $w_{A}$ and the state
$A$ is characterized by the fact that the $a^{\rm th}$ system stay in the state $\alpha_{a}$ and the states $\alpha$ are
the same as at $\eps=0$.  By changing the variables $K$ we change the weights $w_{A}$.  We can thus write
\begin{equation}
q_{t}=\sum_{A,C}w_{A}w_{C}{\sum_{a=1,M} q_{\alpha_{a},\gamma_{a}}\over N}
\end{equation}
It is clear that we have to prove that this way to generate the weights $A$ and $C$ is such the distribution of $q_{t}$
satisfies eq.  \form{CON} .  This will be shown in the next section.

\subsection{Two explicit cases} \label{TWO}
Let us consider here the previous construction in two  cases where we can perform the relevant computations in an 
explicit way. 

The fist case we consider is one step replica symmetry breaking. Here the system may stay in states labeled by
an index $\alpha$. Each state is characterized by a free energy $f_{\alpha}$ and the probability of finding a state in 
the interval $[f,f+df]$ is given by
\begin{equation}
{\cal N}(f)=\exp(m\beta (f-f_{0})) \ .\label{ONESTEP}
\end{equation}
The states of the Gibbsian ensemble will be characterized by the weights $w_{A}$. These states will have the same 
distribution of probability (eq. (\ref{ONESTEP})). However they will be characterized by variables $\alpha_{a}$ that are 
different for each state. Therefore we can write:
\begin{equation}
q_{i}^{A,C}= {\sum_{a=1,M} \sigma_{i}^{\alpha_{a}} \sigma_{i}^{\gamma_{a}} \over N}
\end{equation}
for $A\ne C$ and 
\begin{equation}
q_{i}^{A,A}= {\sum_{a=1,M} \sigma_{i}^{\alpha_{a}} \sigma_{i}^{\alpha_{a}} \over N} \ ,
\end{equation}
where $\sigma_{i}^{\alpha}$ is an equilibrium configuration of the state $\alpha$.

In the limit $M \to \infty$ we can use the law of large numbers and we find with probability one that
\bea
q_{i}^{A,C}=\overline{q_{i}^{\alpha,\gamma}} \equiv q^{0}_{i} \ , \\
q_{i}^{A,A}=\overline{q_{i}^{\alpha,\alpha}} \equiv q^{1}_{i} \ ,
\eea
where the bar denotes the average over all the states, we have used the notation:
\begin{equation}
q_{i}^{\alpha,\gamma}=m_{i}^{\alpha}m_{i}^{\gamma}
\end{equation}
and $m_{i}^{\alpha}$ the magnetization at the site $i$ in the state $\alpha$: 
\begin{equation}
m_{i}^{\alpha} \equiv \lan \sigma_{i}\ran \ .
\end{equation}
We finally find that 
\begin{equation}
P_{i}(q)=m \delta(q-q^{0}_{i})+(1-m)\delta(q-q^{1}_{i}) \ .
\end{equation}
In other words the construction we have used performs automatically the average over all the possible states. It is 
remarkable that in this case the value of $m$ is constant all over the system, and the site variability is present
only in the values of $q^{0}_{i}$ and $q^{1}_{i}$.

The same computations can be done in the case where replica symmetry is broken at two steps.  In this case the states
can be clustered into families, each state is labeled by two indices (e.g. $c,\gamma$) the first index label the family
and the second index labels the state into the given family.  In this case we have two free energy distributions of the
form given in \form{ONESTEP} , one for the states (characterized by a parameter $m_{s}$) and the second for the families
(characterized by a parameter $m_{f}$), where $m_{s}>m_{f}$.  In the same way as before we find
\begin{equation}
P_{i}(q)=m_{f} \delta(q-q^{0}_{i})+(m_{s}-m_{f})\delta(q-q^{1}_{i})+(1-m_{s})\delta(q-q^{2}_{i}) \ ,
\end{equation}
where
\be
\overline{q_{i}^{c,\gamma;d,\delta}} = q^{0}_{i} \ ,\ \ \ \ \
\overline{q_{i}^{c,\gamma;c,\delta}} = q^{1}_{i}\ ,\ \ \ \ \ \
\overline{q_{i}^{c,\gamma;dc\gamma}} = q^{2}_{i}\  .
\ee


We notice that in this approach we have two quantities: the weights, that are global quantities, and the magnetizations 
that depend on the point. For simplicity let us restrict our analysis   to the one step replica symmetry breaking.
In this case for each point we 
can reconstruct the probability distribution of $q_{i}$ if we know the probability distribution $dP_{i}(m)$ of finding 
a magnetization $m_{i}$ at the site $i$ in a generic state:
\be
q_{i}^{0}=\left(\int dP_{i}(m) m\right)^{2}\ ,\ \ \ \ \
q_{i}^{1}=\int dP_{i}(m) m^{2}
\ee

It may be interesting to note that on the Bethe lattice, in the case where one step replica symmetry breaking is exact,
the probability distribution $P_{i}(m)$ depends only on the local environment (the coupling of the nearby points) and it
 can be computed in the large $N$ limit by solving local equations \cite{BETHE,MEPAZE}, confirming the fact, that will be
proved in full generality in the next section, that also the probability distribution $P_{i}(q)$ depends on the local
environment.

\section{The Global Fluctuation Dissipation Relations} 
In this section we will find a short summary of the main results concerning the global FDR.

The usual equilibrium fluctuation theorem can be formulated as follows.  If we consider a pair of conjugated variables
(e.g. the magnetic field and the magnetization) the response function and the spontaneous fluctuations of the
magnetization are deeply related.  Indeed if $R_{eq}(t)$ is the integrated response (i.e. the variation of the
magnetization at time $t$ when we add a a magnetic field from time 0 on) and $C_{eq}(t)$ is the correlation among the
magnetization at time zero and at time $t$, we have that $ R_{eq}(t)=\beta (C_{eq}(0) -C_{eq}(t)) $, where $\beta =
(kT)^{-1}$ and $3/2\ k$ is the Boltzmann-Drude constant.  If we we eliminate the time and we plot parametrically
$R_{eq}$ as function of $C_{eq}$ we have that
\begin{equation}
-{ d R_{eq} \over dC_{eq} }=\beta
\end{equation}
The previous relation can be considered as the definition of the temperature and it is a consequence of the zeroth law
of the thermodynamics.

The generalized fluctuation dissipation relations (FDR) can be formulated as follows in an 
aging system.  Let us suppose that the system is carried from high temperature to low 
temperature at time 0 and it is in an aging regime.  We can define a response function 
$R(t_{w},t)$ as the variation of the magnetization at time $t$ when we add a a magnetic 
field from time $t_{w}$ on, in a similar way $C(t_{w},t)$ is the correlation among the 
magnetization at time $t_{w}$ and at time $t$.  We can define a function $R_{t_{w}}(C)$ if 
we plot $R(t_{w},t)$ versus $C(t_{w},t)$ by eliminating the time $t$ (in the region 
$t>t_{w}$ where the response function is different from zero.The FDR state that for large 
$t_{w}$ the function $R_{t_{w}}(C)$ converge to a limiting function $R(C)$.  We can 
define
\begin{equation}
-{ d R \over dC} =\beta X(C)
\end{equation}
where $X(C)=1$ for $C>C_{\infty}=\lim_{\to \infty}C_{eq}(t)$, and $X(C)<1$ for 
$C<C_{\infty}$.  
The shape of the function $X(C)$ give important information on the free energy landscape 
of the problem, as discussed at lengthy in the literature.  

It has been shown that in stochastically stable system the function $X(C)$ is related to basic equilibrium properties of
the system.  Let us illustrate this point by considering for definitiveness the case of a spin glass.

Spin glasses are characterized by the presence of a quenched disorder (i.e. the coupling $J$ among spins).  For a given
probability distribution of the $J$, there are quantities that do not depend on the particular (generic) realization of
the $J$ and are called self-averaging: the response function and the correlation of the total magnetization belong to
this category.  On the contrary there are other quantities that depend on the choice of the $J$.  A typical example of
non self-averaging quantity is given by the probability distribution of the overlap.  For fixed value of $J$, given two
equilibrium configurations $\sigma$ and $\tau$, we define the global overlap as
\begin{equation}
q(\sigma,\tau) ={\sum_{i=1,N}\sigma_{i}\tau_{i} \over N}
\end{equation}
where $N$ is the total number of spins.  Let us suppose that there is a magnetic field 
(albeit infinitesimal) in such a way that the overlap is positive, otherwise the overlap 
is usually defined as the absolute value of the previous expression.

The probability of  distribution of $q$ is $P_{J}(q)$ and it depends on $J$. The function $P(q)$ is 
defined as the average of $P_{J}(q)$ over the  difference choices of the coupling $J$ and 
obviously depends of the probability distribution of the variable $J$.

It is convenient to introduce the function $x(q)$ defined as
\begin{equation}
x(q)=\int_{0}^{q}P(q')dq'
\end{equation}
or equivalently
\begin{equation}
{d P(q) \over d q }= x(q)
\end{equation}
Obviously $x(q)=1$ in the region where $q>q_{EA}$, where $q_{EA}$ is the maximum value of $q$
where the $P(q)$ is different from zero.

The announced relation among the dynamic FDR and the statics quantities is simple
\begin{equation}
X(C)=x(C)
\end{equation}
We shall see later that this basic relation  can be derived using the principle of stochastic stability that
assert that the thermodynamic properties of the system do not change too much if we add a 
random perturbation to the Hamiltonian.  All that is well known.  In the nest section we 
shall se how to play the same music will local variables.
\section{The Local Fluctuation Dissipation Relations} 

There are recent results that indicate that the FDR relation and the static-dynamics 
connection can be generalized to local variables in systems where a quenched disorder is 
present and aging is heterogeneous, We shall see that there finding need a more general framework to be 
explained. 

For {\sl one given sample} we can consider the local integrated response function 
$R_{i}(t_{w},t)$, that is the variation of the magnetization at time $t$ when we add a 
magnetic field at the point $i$ starting at the time $t_{w}$.  In a similar way the local 
correlation function $C_{i}(t_{w},t)$ is defined to the correlation of among the spin at 
the point $i$ at different times ($t_{w}$ and $t$).  Quite often in system with quenched 
disorder aging is very heterogenous: the function $C_{i}$ and $R_{i}$ change dramatically 
from on point to an other.  

It has been observed in simulations that local fluctuation dissipation relations (LFDT) seems to hold
\begin{equation}
-{ d R_{i} \over dC_{i} } =\beta X_{i}(C)\ ,
\end{equation}
where $X_{i}(C)$ has quite strong variations with the site.

It  has also been suggested that in spite of these strong heterogeneity, if we define the effective 
$\beta^{eff}_{i}$ at time $t$ at the site $i$ as ,
\begin{equation}
-{ d R_{i}(t_{w},t) \over dC_{i}(t_{w},t) } =\beta X_{i}(t_{w},t)\equiv\beta^{eff}_{i} 
(t_{w},t) \ ,
\end{equation}
the quantity $\beta^{eff}_{i}$ does not depend on the site.  In other words a thermometer 
coupled to a given site would measure (at a given time) the same temperature independently 
on the site: different sites are thermometrically indistinguishable.

These empirical results calls for a theoretical explanation.  The aim of this note is to 
show that these results are consequence of stochastic stability in an appropriate contest 
and that there is a local relation among static and dynamics.  In the next section we will 
define in an appropriate way the local probability distribution of the overlap for a given 
system at point $i$ (i.e. $P_{i}(q)$). We will also define the function $x_{i}(q)$  as
\begin{equation}
x_{i}(q)=\int_{0}^{q}P_{i}(q')dq' \ 
\end{equation}
and we will show that the static-dynamic connection for local variables is very similar to 
the one for global variables and it is given by:
\begin{equation}
X_{i}(C)=x_{i}(C)
\end{equation}

In order to arrive to to prove of the local FDR, that will be presented in the next section, is it convenient to
reconsider the dynamic definition of the correlation function $C_{i}(t_{w},t)$.  It is clear that $C_{i}(t_{w},t)$
cannot be measured by observing only one single history of our sample: $\sigma_{i}(t_{w}) \sigma_{i}(t_{w}+t) =\pm 1$. 
The correlation function is obtained by averaging over all the possible history, i.e. by repeating the experiment $M$
times and sending eventually $M$ to infinity.  In other words the two time local correlation function is not a self
averaging quantity as far histories are concerned.

If we want to define the correlation in such a way that it can be measured by observing a 
single history we have to consider the Gibbsian introduced in section II.  We can consider $M$ identical 
copies (or clones) of our sample and the Hamiltonian in this Gibbs ensemble is just given
by
\begin{equation}
H_{0}(\sigma)=\sum_{a=1,M} H(\sigma^{a}) \ .
\end{equation}

Obviously for large $M$ the correlation function can be defined as
\begin{equation}
C_{i}(t_{w},t)={\sum_{a=1,M} \sigma_{i}^{a}(t_{w}) \sigma_{i}^{a}(t) \over M}
\end{equation}
and in the limit $M\to\infty$ of this quantity is self-averaging, i.e. it will be the same in
all the history of the system.  It is evident that the $M$ systems are independent so that
the average of one Gibbsian systems is equivalent to the average of $M$ usual systems and
corresponds to repeat the same experiment (or computer run) $M$ times.  A similar
procedure can be done for the response function.

\section{Perturbing the system}
\subsection{Global perturbation}
As usual we can rederive the probability distribution of the overlaps by perturbing the system.
Let us consider for simplicity the effect of adding to the Hamiltonian $H_{K}(\sigma)$ an extra 
perturbation given by
\begin{equation}
\Delta H^{(1)} \equiv \sum_{i=1,N,a=1,M} h^{a}_{i}\sigma^{a}_{i} \ ,
\end{equation}
where the variables $h$ are Gaussian random variables with zero average and variance $\delta$.

By integration by part we find that for given $K$
\begin{equation}
{\lan \Delta H^{(1)} \ran \over NM} =
{\delta  \beta \sum_{i=1,N,a=1,M} (1 -\lan  \sigma^{a}_{i} \ran ^{2}) \over NM}=
\delta \beta \int dq P_{t}(q) (1-q) \ .
\end{equation}
Therefore we have the relation
\begin{equation}
\chi^{(1)}\equiv \lim_{\delta\to 0}{\lan \Delta H^{(1)} \ran \over \delta NM }=\beta \int dq P_{t}(q) (1-q)
\end{equation}
 for the susceptibility $\chi^{1}$ . The proof is identical to the standard one. On the other hand 
 at $\eps=0$ stochastic stability implies that 
 \begin{equation}
 \chi^{(1)}=\beta \int dq P(q) (1-q) \ ,
 \end{equation}
 where $P(q)$ is the the usual $J$ average of the $J$ dependent probability distribution.
 
 It is a trivial task to generalize the proof to the other susceptibilities. For example if we define
 \begin{equation}
\Delta H^{(2)} \equiv \sum_{i=1,N,a=1,M,k=1,N,b=1,M} h^{a,b}_{i,k}\sigma^{a}_{i} \sigma^{b}_{k} \ ,
\end{equation}
where the variables $h$ are Gaussian random variables with zero average and variance $\delta/(NM)$,
we get 
\begin{equation}
\chi^{(2)}\equiv \lim_{\delta\to 0}{\lan \Delta H^{2} \ran \over \delta NM} =\beta \int dq P_{t}(q) 
(1-q^{2})=
\beta \int dq P_{t}(q) (1-q^{2}) \ .
\end{equation}

In this way we can compute all the moments of the both the function  $P_{t}(q)$ and $P(q)$ and they 
coincide. The two functions are equal.

\subsection{Local perturbation}
We can now repeat the same steps as before but locally.
Let us consider for simplicity the effect of adding to the Hamiltonian $H_{K}(\sigma)$ an extra 
perturbation given by
\begin{equation}
\Delta H^{(1)}_{i} \equiv \sum_{a=1,M} h^{a}_{i}\sigma^{a}_{i} \ ,
\end{equation}
where the variables $h$ are Gaussian random variables with zero average and variance $\delta$.
By integration by part we find that for given $K$
\begin{equation}
{\lan \Delta H^{(1)}_{i} \ran \over NM} =
{\delta  \beta \sum_{a=1,M} (1 -\lan  \sigma^{a}_{i} \ran ^{2}) \over M}=
\delta \beta \int dq P_{i}(q) (1-q) \ .
\end{equation}
Therefore we have the relation
\begin{equation}
\chi^{(1)}_{i}\equiv \lim_{\delta\to 0}{\lan \Delta H^{(1)}_{i} \ran \over \delta M} =\beta \int dq P_{i}(q) 
(1-q)
\end{equation}
 for the susceptibility $\chi^{(1)}_{i}$ .

 Let us consider an aging system and add the perturbation $\Delta H^{(1)}_{i}$ at time $t_{w}$.
 We have that
\begin{equation}
\chi^{(1)}_{i}(t)=\lim_{\delta\to 0}\lan \Delta H^{(1)}_{i} \ran_{t}={\sum 
_{a=1,M}R_{i,}^{a}(t_{w},t) \over M}= 
\end{equation}
However in the limit  small $\eps$ in the dynamics does not depends on $a$ so that we get
\begin{equation}
\chi^{(1)}_{i}(t)=R_{i,}(t_{w},t) \ .
\end{equation}
Assuming that 
\begin{equation}
\lim_{t \to \infty}\chi^{(1)}_{i}(t)=\chi^{(1)}_{i} \ ,
\end{equation}
we arrive to 
\begin{equation}
\lim_{t \to \infty}R_{i,}(t_{w},t')= \chi^{(1)}_{i}=\beta \int dq P_{i}(q) (1-q) \ .
\end{equation}

We can now copy {\sl mutatis mutandis} the proof of the usual FDR. For example let us  define
 \begin{equation}
\Delta H^{(2)}_{i} \equiv \sum_{a=1,M,b=1,M} h^{a,b}\sigma^{a}_{i} \sigma^{b}_{i} \ ,
\end{equation}
where the variables $h$ are Gaussian random variables with zero average and variance $\delta/(M)$. 
The static susceptibility is given by
\begin{equation}
{\lan \Delta H^{(2)}_{i} \ran \over NM} =
\delta  \beta \sum_{a,b=1,M} {1 -\lan  \sigma^{a}_{i}  \sigma^{b}_{i} \ran ^{2} ) \over M}=
\delta \beta \int dq P_{i}(q) (1-q^{2}) \ . \label{STATICHI}
\end{equation}

If we assume for simplicity a Langevin type of evolution, the same steps of \cite{NOI} give
we have that
\begin{equation}
\chi^{(2)}_{i}(t)=2\int_{t_{w}}^{t}d t'C_{i}(t',t) {\partial R_{i}(t',t) \over \partial t'} \ .
\end{equation}
Assuming for simplicity the asymptotic scaling (this step is not necessary indeed the following 
formulae force its validity) we get
\begin{equation}
\lim_{t\to \infty}\chi^{(2)}_{i}(t) =2 \int dC C\ X_{i}(C) \ ,
\end{equation}
where we have defined:
\begin{equation}
X_{i}(t)(C(t',t))={\partial R_{i}(t',t) \over \partial t'}\left({\partial C_{i}(t',t) \over \partial t'}\right)^{-1}
\end{equation}
and 
\begin{equation}
X_{i}(C)=\lim_{t \to \infty} X_{i}(t)(C) \ .
\end{equation}
If also in this case the limits $t\to \infty $ and $\delta \to 0$ may be exchanged we get
\begin{equation}
\lim_{t \to \infty}\chi^{(2)}_{i}(t)=\chi^{(2)}_{i}=\beta \int dq P_{i}(q) (1-q^{2})\ .
\end{equation}

Generalizing the previous arguments we get
\bea
\chi^{(s)}_{i}(t)=s\int_{t_{w}}^{t}d t'(C_{i}(t',t)^{s-1}){\partial R_{i}(t',t) \over \partial t'}=
s \int dC C^{s-1}X_{i}(C)= \\
\beta \int dq P_{i}(q) (1-q^{s})=s \beta\int x_{i}(q)q^{s-1}\ .
\eea
We thus arrive to  the announced conclusion that 
\begin{equation}
X_{i}(C)=x_{i}(C)\equiv\int_{0}^{C} dqP_{i}(q) dq\ .
\end{equation}
This is the local relation among the static and the local fluctuation dissipation relations.

A few remarks are in order
\begin{itemize}
    \item If we take a sequence of larger and larger systems, the dynamical quantities converge to a well defined limit
    when the volume goes to infinity.  Therefore also the local overlap distribution $P_{i}(q)$ goes to a limit when the
    volume goes to infinity and depends only an the local environment (i.e. the $J$ not too far from the point $i$).  It
    is clear that all the problems due to the definition of the function $P_{J}(q)$ infinite volume limit due to the
    difficulties of define the infinite volume limit of local observation fades away.  It is quite possible that the
    construction we have presented in this paper may be useful in a rigorous approach.

    \item The way in which we have coupled together our $M$ clones is irrelevant.  The only thing we need to introduce
    such a coupling is to correlate the states of one clones with the states of the the other clone.  
\end {itemize}

\subsection{Bilocal Perturbations}
We are now ready for proving thermometric indistinguishability of the sites.  We consider two far away sites $i$ and $k$
and we perturbation that depends both the spins at $i$ and the spins at $k$.

A typical example is
\begin{equation}
\Delta H^{(3,2)}_{i,k} \equiv \sum_{a_{1},a_{2},a_{3},b_{1},b_{2}=1,M} h^{a_{1},a_{2},a_{3},b_{1},b_{2}}
\sigma^{a_{1}}_{i} \sigma^{a_{2}}_{i} \sigma^{a_{3}}_{i} \sigma^{b_{1}}_{i} \sigma^{b_{2}}_{i} \ .
 \end{equation}
where the variables $h$ are Gaussian random variables with zero average and variance $\delta/M^{4}$. 
In the same way as before we get that
\begin{equation}
\chi^{(s_{i},s_{k})}_{i,k}=\beta \int dq_{i}dq_{k}P(q_{i},q_{k}) (1-(q_{i})^{s_{i}}(q_{k})^{s_{k}})\ ,
\end{equation}
where $P(q_{i},q_{k})$ is the probability distribution of  $q_{i}$ and $q_{k}$.
If we compute the same quantity in the aging regime for very large time we get that the same 
quantity must be equal to the large time limit of 
\begin{equation}
\int _{t_{w}}^{t}dt \[[ X_{i}(t){\partial C_{i}^{s_{i}}\over \partial t} C_{k}^{s_{k} }+
X_{k}(t)C_{i}^{s_{i}} {\partial C_{k}^{s_{k}}\over \partial t}\]] \ .
\end{equation}
Imposing that the two expressions can be equal for arbitrary $s_{i}$ and $s_{k}$ one 
recover the form of the $P(q_{i},q_{k})$ in terms of $X_{i}(t)$ and $X_{k}(t)$. By 
imposing that $P(q_{i},q_{k})$ is positive (i.e. it does not contain a derivative of 
a delta function) one finds that
\begin{equation}
X_{i}(t)=X_{k}(t) \ . \label{PRIMA}
\end{equation}
We finally obtain
\begin{equation}
P(q_{i},q_{k})=\int_{0}^{1}dx\delta(q_{i}-q_{i}(x))\delta(q_{k}-q_{k}(x))\ , \label{SECONDA}
\end{equation}
where $q_{i}(x)$ is the inverse function of $x_{i}(q)$.

The first conditions is just thermometric indistinguishability of the sites during aging while the 
second condition has some interesting consequences that will be investigated elsewhere.

We note that the probability distribution of the local overlap $P_{i}(q)$, being related to a dynamical quantity,  must 
depend only on the local environment around the point $i$ and therefore it must have a straightforward limit when the 
volume of the system goes to infinity (e.g it should be independent of the boundary conditions). It is remarkable that for 
far away points ($i,k$) the two probability distributions   $P_{i}(q)$ and $P_{k}(q)$ are independent one from the 
other, but the joint probability distribution of $q_{i}$ and $q_{k}$ does not factorize has shown by eq. \ref{SECONDA}. 

\section{Computing the local overlap distribution}
The formulae presented in the previous section are useful to define the local overlap and its distribution, but they are
not handy as far as practical computations are concerned.  The distribution $P_{i}(q)$ will have rounded delta functions,
 for finite $M$.  On the other hand the burden of the numerical computation increase violently with $M$.  The
best thermalization method (the parallel tempering) becomes more and more slow when the volume of the system ($NM$)
increases. It is convenient to obtain this kind of information using different methods.

\subsection{Computing the moments}

In many cases a the knowledge of the first few moments of the function $P_{i}(q)$ is enough to compute this function. 
This approach is very useful in the case where we have some a priori information on the shape of this function. For 
example in the case of one step broken replica symmetry the knowledge  of the first two moments and of $m$ 
determines completely the function $P_{i}(q)$. 

In other models, where the replica symmetry is broken in a continuos 
way it is possible that a good approximation is given by simple expressions like
\be
P_{i}(q)=\theta(q-q^{EA}_{i})(1-q)^{-1/2}Q_{i}(q)+m\delta(q-q^{EA}_{i})
\ee
where $Q_{i}(q)$ is a low degree polynomial. However the validity of similar formulae may strongly change from model to 
model.

The computation of the moments is rather simple. We consider the Gibbsian system  with fixed $M$. If  $\sigma$ and 
$\tau$ are two equilibrium configurations of the model it is possible to prove that 
\be
\overline{\lan \prod_{a=1,M}\sigma^{a}_{i}\tau^{a}_{i} \ran} =q^{(M)}_{i}\equiv \int dq \ q^{M} P_{i}(q) \ ,
\ee
or more generally
\be
\overline{\lan \sigma^{a_{1}}_{i}\tau^{a_{1}}_{i}\ldots\sigma^{a_{k}}_{i}\tau^{a_{k}}_{i}  \ran} =q^{(k)}_{i} \ .
\ee
where the indices  $a$ are arbitrary as soon as they are all different (obviously $k<M$).

The proof of the previous relation can be obtained by computing the local susceptibilities like eq. (\ref{STATICHI}).  Let 
us consider the case $k=2$. We could also define
 \begin{equation}
\Delta H^{(2)}_{i} =\left({M \over M-1}\right)^{1/2} \sum_{a=1,M,b=1,M;a\neq b} h^{a,b}\sigma^{a}_{i} \sigma^{b}_{i} \ ,
\end{equation}
The corresponding susceptibility should not depend on $M$ and it is trivial equal to  the previous defined susceptibility 
when $M\to \infty$. In other words 
\be
\overline{\lan \sigma^{a}_{i}\tau^{a}_{i}\sigma^{b}_{i}\tau^{b}_{i}  \ran}  \ .
\ee
does not depend on $M$, $a$ and $b$ as soon as $a\ne b$.

This result can be checked in an explicit way in the two explicit cases discussed before \form{TWO} .

The computation of the moment of order $k$ involves only  the thermalization of a system with  only $k N$ sites and it
can be done with not too much computational effort for not too large $k$.  

\subsection{Introducing a state reservoir}
An other possibility, that is more interesting especially for analytic computations, consists in noting that the local
overlap distribution depends only on the local environment.  Therefore we can embed the local environment in a large
system, and the local properties should not be related to that of a rest of the system as soon as the rest of the system
has the same distribution of states of the original system.

Let us consider a simple case, a spin glass model in three dimensions. Let us suppose that within the needed accuracy
the knowledge of the couplings up to a distance $R$ from the site $i$ determines the function $P_{i}(q)$.  Let us
consider a system of size $L>4R$ with the same probability distribution of the couplings, with the constraint however
that  there are two points $i_{1}$ and $i_{2}$ such that the local environments of radius $R$ around each of these two
points (that are at distance greater than $2R$) are the same and equal to that of the original system at the point $i$. 

In this case the same argument as before leads to the conclusions that if we take two different equilibrium
configurations $\sigma$ and $\tau$, we have that
\be
\overline{\lan \sigma_{i_{1}} \sigma_{i_{2}} \tau_{i_{1}} \tau_{i_{2}}\ran} =q^{(2)}_{i} \ ,
\ee
where the bar denotes the average over the couplings that do not belong to the fixed environment.

A different possibility would be to take two copies of the local environment and to couple the spins at the boundary of
each copy to other spins  that stay on a Bethe lattice. If the state distribution of the Bethe lattice is the same as 
that of the three dimensional lattice, one should get identically results.

In both case the main role of the system outside the local environment was to stay in different possible states, with 
the correct probability distribution. In the nutshell played the role of a state reservoir. As far as the detailed 
nature of this reservoir is not important we can consider the simplest possible model for it. A very convenient choice 
is the following:
we model the interaction of the local 
environment with the rest of the world by introducing fields an interaction
\be
\sum_{k\in S} h_{k}\sigma_{k} \ ,
\ee
where the sum is done over the spins of the surface ($S$). The free energy corresponding to a given choice of the variables 
$h$ is $F[h]$ and the corresponding magnetization of the site $i$ is $m[h]$. 

We should now modeling the ensemble of the $h$ and of the $w$.  We should introduce the distribution of the weight
$w_{\alpha}$ and for each state $\alpha$ we should give the set of the $h$.  Let us consider the simplest case of a
system with one step replica symmetry where we fix the value of $m$.  In this case we can assume that the $w$ are
distributed according to the expression in \form{ONESTEP} .  We can assume the variable $h^{\alpha}$ have a given
probability distribution, e.g. one of the simplest possibility is that
\begin{equation}
h^{\alpha}_{k}= \overline{h}_{\alpha}+\delta h^{\alpha}_{k}
\end{equation}
where  both $\overline{h}_{\alpha}$ and $\delta h^{\alpha}_{k}$ are uncorrelated random variables with zero average and 
variance $h^{(0)}$ and $h^{(1)}$. Different form of the probability distribution of the fields $h$ can be used, however
if the local environment is sufficient large, the result should not depend on the form of this probability distribution 
or from the variances $h^{(0)}$ and $h^{(1)}$. 

One finally finds that the two parameters that  identify the probability distribution, i.e. $q^{0}_{i}$ and
$q^{1}_{i}$ are given respectively
\bea
q^{0}_{i}=
\left( 
\overline{\sum _{\alpha=1,{\cal A}} w_{\alpha} \exp(-\beta F[h^{\alpha}]) m[h^{\alpha}] \over
\sum _{\alpha=1,{\cal A}} w_{\alpha} \exp(-\beta F[h^{\alpha}])}
\right)^{2}\\
q^{1}_{i}
=\overline{\sum _{\alpha=1,{\cal A}} w_{\alpha} \exp(-\beta F[h^{\alpha}]) m[h^{\alpha}]^{2} \over 
\sum _{\alpha=1,{\cal A}} w_{\alpha} \exp(-\beta F[h^{\alpha}])}
\eea
where $\cal A$ is a large number, the bar denotes the average over the magnetic field $h$ and the weights $w$.
Using the techniques of \cite{BETHE}
in the one step replica breaking case the average over the $w$ can be done and one obtains;
\bea
q^{0}_{i}=
\left( 
\overline{\sum _{\alpha=1,{\cal A}} \exp(-\beta m F[h^{\alpha}]) m[h^{\alpha}] \over
\sum _{\alpha=1,{\cal A}}  \exp(-\beta m F[h^{\alpha}])}
\right)^{2}\\
q^{1}_{i}
=\overline{\sum _{\alpha=1,{\cal A}}  \exp(-\beta m F[h^{\alpha}]) m[h^{\alpha}]^{2} \over 
\sum _{\alpha=1,{\cal A}}  \exp(-\beta m F[h^{\alpha}])}
\eea

Of course if our model is a Bethe lattice, the computations are quite simple and we reobtain the results of \cite{BETHE}, 
i.e. a local version of the equations of \cite{MEPAZE}.

\section{Conclusions}
The main results of this note is the definition of a local probability distribution of the overlap, that depend on the
site for fixed sample.  The properties of this local probability distribution are related to the local fluctuation
dissipation relations, that automatical follows from the present formalism.  The property of thermometric
indistinguishability of the sites turns out to be a byproduct of our approach: during an aging regime all the sites are
characterized by the same effective temperature during the aging regime. 

The two times local correlations functions can be written
\be
C_{i,}(t_{w},t)=C_{i}(x(t_{w},t))
\ee
where the function $C_{i}(x)$ is the inverse of the function $x_{i}(C)$ and can be obtained only by static 
measurements. The whole local dependence of off-equilibrium correlations and responses can be computed from static 
quantities. The only quantity that cannot be computed from equilibrium consideration is the global effective 
temperature, i.e. 
$\beta x(t_{w},t)$.

The physical picture that emerges is quite clear. The local overlap distribution can be defined by averaging over the 
ensemble of the states of the system and the introduction of the Gibbs ensemble is a technical tool to perform this  
average in a constructive way. During the dynamics the system locally explore different states of the system in a 
random way. It should not be surprising that the dynamical average is deeply connect to the static average on all 
possible different states.

I am grateful to Jorge Kurchan, Marc M\'ezard and Federico Ricci-Tersenghi for  illuminating discussions.
\section*{Appendix}
It is interesting to consider what happens if we look to the model with Hamiltonian 
\begin{equation}
\sum_{a=1,M} H(\sigma^{a})  +\eps \sum_{a,b=1,M;i,j=1,N}K^{a,b}_{i,j}\sigma_{i}^{a}\sigma_{j}^{b}
\end{equation}
in the limit $M \to \infty$ but with fixed $N$..
In this case the second part of the Hamiltonian coincide with that of the Sherrington-Kirkpatrick model.

For small enough $\eps$ one finds that the solution of the model is the replica symmetric one. It depends on a 
parameter $q$, that can be found by looking to the solution of the equations:
\be
q={\sum _{i=1,N}\overline{m_{i}[h]^{2}}\over N}\ ,
\ee
where the magnetization are obtained by considering the statistical average for one sample with Hamiltonian
\be
H(\sigma)+\sum_{i}h_{i}\sigma_{i}
\ee
and the magnetic fields are  random independent identically distributed Gaussian variables with zero average and
variance $\eps q$ and the bar denotes the average over the fields $h$.

However by increasing $\eps$ this solution may becomes unstable (De Almeida Touless line) and at higher $\\esp$ one has
to look for replica broken solutions.  A computation
similar to the original one  shows that
%
replica broken phase happens when $\eps^{2} N=O(1)$ and this result is at the origine of the condition $\eps^{2 }N >>1$.


\begin{thebibliography}{99}
 \bibitem{RM} E. Marinari, G. Parisi, F. Ricci-Tersenghi and J. J. Ruiz-Lorenzo, J. Phys.  A: Math and Gen.  {\bf 31},
L481 (1998); E. Marinari, G. Parisi and J. Ruiz-Lorenzo, {\em Numerical Simulations of Spin Glass Systems}, in {\em Spin
Glasses and Random Fields}, edited by P. Young (World Scientific, Singapore 1998); E. Marinari, G. Parisi, F.
Ricci-Tersenghi, J. Ruiz-Lorenzo and F. Zuliani, J.Stat. Phys. {\bf 98},  973 (2000).
  
\bibitem{I} G.Parisi {\em Local fluctuation dissipation relations} cond/mat

\bibitem{CUKU} L.F. Cugliandolo and J. Kurchan, { Phys.  Rev.  Lett.} { 71}, 173 (1993); { J. 
Phys.  A: Math.  Gen.} {\ 27}, 5749 (1994).    

\bibitem{framez}S. Franz and M. M\'ezard, Europhys.  Lett.  {\bf 26}, 209 (1994).

\bibitem{CKPA} L.F. Cugliandolo, J. Kurchan and G.  Parisi, J. Phys. I (France) {\bf 4}, 1641 (1994).

\bibitem{revaging} For a review see J.P.~Bouchaud, L.F.~Cugliandolo, J.~Kurchan and M.~M\'ezard, {\it Out of equilibrium
dynamics in spin glasses and the other glassy systems, } in {\it Spin Glasses and Random Fields, } edited by P.~Young,
World Scientific (1997).

\bi{GUERRA} F.  Guerra, Int.  J.  Phys.  B, {\bf 10}, 1675 (1997).

\bi{AI} M. Aizenman and P. Contucci, J. Stat.  Phys.  {\bf 92} 765 (1998).

\bi{SOL} G.  Parisi, {\sl On the probabilistic formulation of the replica approach to spin glasses},
 cond-mat/9801081.
 
\bibitem{NOI}S. Franz, M. M\'ezard, G. Parisi, L. Peliti, Phys.  Rev.  Lett.  {\bf 81}, 1758 (1998), J. Stat.  Phys. 
{\bf 97} 459 (1999) .

\bibitem{FR} S. Franz and H. Rieger Phys.  J. Stat. Phys.  {\bf 79} 749 (1995).

\bibitem{P} G. Parisi Phys. Rev. Lett. {\bf 78} 4581  (1997).

\bibitem{K} W. Kob and J.-L. Barrat, Phys. Rev. Lett. {\bf 79} 3660 (1997).

\bibitem{IG}T. S. Grigera and N. E. Israeloff, PRL {\bf 83}, 5038 (1999).

\bibitem{BC} Bellon and S. Ciliberto, cond-mat/0201224, to be published in Physica D (2002).

\bibitem{HO}  D. H\'{e}risson and M. Ocio, Phys. Rev. Lett.  Letters \textbf{88}, 257202 (2002)


\bibitem{CC} H.E.~Castillo, C. Chamon, L. Cugliandolo and M. Kennett, Phys.  Rev.  Lett.  \textbf{88}, 237201 (2002).

\bibitem{MR} A. Montanari and F. Ricci-Tersenghi, {\sl A microscopic description of the aging dynamics:
fluctuation-dissipation relations, effective temperature and heterogeneities}, cond-mat 0207416.

\bibitem{CKP}L. Cugliandolo, J. Kurchan and L. Peliti, Phys.  Rev.  {\bf E55} 3898 (1997).

\bibitem{FV}  S. Franz and M.A. Virasoro, J. Phys.  A: Math.  and Gen.  {\bf 33}, 891 (2000).


\bibitem{HE} P.H.~Poole {\it et al.}, Phys.  Rev.  Lett.  \textbf{78}, 3394 (1997);  A.~Barrat and R.~Zecchina, Phys. 
Rev.  E \textbf{59}, R1299 (1999);  F.~Ricci-Tersenghi and R.~Zecchina, Phys.  Rev.  E \textbf{62}, R7567 (2000).


\bibitem{BETHE} M. M\'ezard and G. Parisi

\bibitem{MEPAZE} M.~M\'ezard, G.~Parisi and R.~Zecchina, Science{\it xpress} 10.1126/science.1073287 (2002). 
M.~M\'ezard and R.~Zec\-chi\-na, {\tt cond-mat/0207194}.

\end{thebibliography}
\end{document}